\documentclass[final,5p,times,twocolumn]{elsarticle}

\usepackage{amssymb}

\journal{Physics Letters B}

\begin{document}

\begin{frontmatter}



\title{Measurement of the vector and tensor analyzing powers for $dp$- elastic scattering at 880~MeV}


\author[LHE]{P.K.Kurilkin}
\address[LHE]{VBLHEP-JINR, 141980 Dubna, Moscow region, Russia}
\author[LHE]{V.P.Ladygin}
\author[CNS]{T.Uesaka}
\address[CNS]{Center for Nuclear Study, University of Tokyo, Tokyo
  113-0033, 
Japan}
\author[RIKEN]{K.Suda}
\address[RIKEN]{RIKEN Nishina Center, Wako, Saitama 351-0198, Japan}
\author[LHE]{Yu.V.Gurchin}
\author[LHE]{A.Yu.Isupov}
\author[Saitama]{K.Itoh}
\address[Saitama]{Department of Physics, Saitama University, Saitama, Japan}
\author[LHE,Zilina]{M.Janek}
\address[Zilina]{Physics Department, University of \v Zilina, 
010 26 \v Zilina, Slovakia}
\author[LHE,Bukh]{J.-T.Karachuk}
\address[Bukh]{Advanced Research Institute for Electrical Engineering, 
Bucharest, Romania}
\author[CNS]{T.Kawabata}
\author[LHE]{A.N.Khrenov}
\author[LHE]{A.S.Kiselev}
\author[LHE]{V.A.Kizka}
\author[LHE,INP]{V.A.Krasnov}
\address[INP]{Institute for  Nuclear Research, Moscow, Russia}
\author[LHE]{N.B.Ladygina}
\author[LHE,INP]{A.N.Livanov}
\author[Kyushi]{Y.Maeda}
\address[Kyushi]{Kyushi University, 6-10-1 Hakozaki, Higashi-ku, 
Fukuoka-shi 812, Japan}
\author[LHE]{A.I.Malakhov}
\author[LHE]{S.M.Piyadin}
\author[LHE]{S.G.Reznikov} 
\author[CNS]{S.Sakaguchi}
\author[TU]{H.Sakai}
\address[TU]{University of Tokyo, Bunkyo, Tokyo 113-0033, Japan}
\author[CNS]{Y.Sasamoto}
\author[RIKEN]{K.Sekiguchi}
\author[LHE]{M.A.Shikhalev}
\author[LHE]{T.A.Vasiliev}
\author[Krakow]{H.Witala}
\address[Krakow]{Institute of Physics, Jagiellonian University,
  Krak\'ow, 
Poland}

\begin{abstract}
The vector $A_y$ and tensor analyzing powers $A_{yy}$ and $A_{xx}$ for $dp$- elastic scattering were measured at $T_d^{lab}$ = 880~MeV over the c.m. angular range from 60$^\circ$ to 140$^\circ$ at the JINR Nuclotron. The data are compared with predictions of different theoretical models based
on the use of nucleon-nucleon forces only. The observed discrepancies of the measured analyzing powers from
the calculations require the consideration of additional mechanisms.

\vspace{1pc}

\end{abstract}


\end{frontmatter}


\section{Introduction}
\label{introduction}

The nucleon-deuteron ($Nd$) elastic scattering is the simplest 
composite-particle scattering process. It has been widely used to 
test how well the few-nucleon systems can be described in terms of 
nucleon-nucleon (NN) interactions.
At energies below the $\pi$ production threshold (E=210~MeV/u), 
Faddeev calculations provide rigorous descriptions of the scattering
process. 
During the last several years, 
polarization observables in $Nd$ elastic scattering
have been studied in a number of experiments at 
RIKEN \cite{Sakamoto,Sakai,Sekiguchi02,Sekiguchi04}, 
KVI \cite{Bieber00,Ermisch01,Ermisch05,Stephan07,Mardan07,Amir07,Ramazani08}, 
IUCF \cite{Stephenson99,Cadman01,Przewoski06} and  
RCNP \cite{Hatanaka02,Maeda07}.
The considerable experimental activities were stimulated, in particular, 
by the observed discrepancy of $\sim$30\% between differential cross 
section data \cite{Sakamoto,Sagara} 
measured at energies of 65--135 MeV/nucleon and results 
of Faddeev calculations \cite{CB0}. Many of the discrepancies 
for the differential cross sections and vector analyzing powers at
these energies are remedied by the inclusion of the 2$\pi$ -exchange 
three-nucleon forces (3NFs) such as TM-3NF \cite{TM},
UrbanaIX-3NF \cite{Urbana9} or TM99 \cite{TM99} into 
the calculations \cite{CB1}. 
But theoretical calculations with 3NFs still
have difficulties in reproducing data of some spin observables, for instance,
tensor analyzing powers.

At high energies, the forward angle $Nd$ elastic 
scattering has been successfully described by the Glauber approach \cite{glauber} 
which takes into account both single and double scattering terms. 
The interference between the single- and double- scattering amplitudes
including the $D$- state in the deuteron wave function (DWF) allowed one 
to explain the filling of the cross section diffractive minimum \cite{glauber-D}.   

%

At energies 200--600 MeV/nucleon, however, large
discrepancies between the experimental 
data \cite{Hatanaka02,Maeda07,Ermisch03,Sekiguchi11} and 
theoretical predictions in the minimum of the differential cross 
section are persistent even after inclusion of 3NFs \cite{CB1,Deltuva}. 
There are several possible origins for these discrepancies. One is 
relativistic effects: the relativistic 
Faddeev approach \cite{Faddeev_rel,Faddeev_rel1} has been 
developed for intermediate energies range. 
It was found that when only NN forces are included in the
calculations, the relativistic effects are
significant only at backward scattering angles. 
They are relatively small in the minimum of the differential cross 
section  where the discrepancies between 2N force predictions 
and data are largest \cite{Faddeev_rel}. The extension of the Faddeev calculation technique
into the relativistic regime \cite{elster3,elster4} also
does not provide a reasonable description of the experimental data, only with NN forces.
 Recently, however, it is reported that large
 changes of the elastic scattering cross section in the region of
 angles ranging from the minimum of the cross section up to very
 backward angles are observed when 3NFs are included in the calculations \cite{Faddeev_rel1}.

Another possibility is the manifestation of new 3NFs. Existing 3NF models \cite{TM,Urbana9,TM99} 
mostly deal with low-momentum diagrams, while diagrams with higher momentum can make 
visible contributions in scatterings at 200--600 MeV/nucleon where large momentum transfers are relevant. 
It is also possible to consider reaction mechanisms which are not included in the present Faddeev calculations. In the energy region considered, effects
due to $\pi$ production and on-shell $\Delta$ excitation are the candidates.

Thus it is strongly anticipated that a reaction theory for 
the nucleon-deuteron scattering at 200--600~MeV is established and 
further information on new 3NF is extracted from the scattering data. 
The energy region lies above the $\pi$- production threshold and the 
applicability of the Faddeev calculation techniques, at least as it
is, is not trivial. On the one hand, these energies are not large enough to
apply the Glauber approach. 

   In this work  new results on the analyzing powers $A_y$, $A_{yy}$ 
and  $A_{xx}$ for $dp$- elastic scattering measured at $T_d^{lab}$=
880 MeV over the c.m. angular range from 60$^\circ$ to
$140^\circ$ are presented and are compared with several theoretical 
calculations, including Faddeev \cite{CB0} and Glauber \cite{glauber} ones. 


\section{Experimental procedure}
\label{experiment}

The experiment has been performed at the 
Internal Target Station (ITS) \cite{ITS} at the superconducting
synchrotron, 
Nuclotron at the Laboratory of High Energy Physics, Joint Institute
for 
Nuclear Research. 
The ITS consists of a spherical scattering chamber and a target sweeping system.
The scattering chamber is fixed on the flanges of the Nuclotron ion tube.
The disk mounting six different targets is 
located on the axle of a stepper motor. A target used for the
measurement 
is moved to the
center of the ion tube when the particles are accelerated up to the 
required energy.
A 10~$\mu$m~CH$_2$ film  was used as a proton target. 
A carbon wire was used to evaluate the background originating from the
carbon content in CH$_2$. 
The signal from the target position monitor \cite{gurchin} was used to
tune the accelerator parameters to bring the interaction point close to the 
center of the ITS chamber. 
The signals from the monitor were also stored as raw data so that one 
can use the position information in off-line analysis.

The polarized deuteron beam was provided by the atomic-beam polarized 
ion source POLARIS
\cite{polaris}. Nuclear polarization is provided via radio-frequency
(RF) 
hyperfine
transitions. In this experiment the data were taken for three spin modes:
unpolarized, "2-6" and "3-5", which have theoretical maximum polarizations of  
$(p_Z,p_{ZZ})$ =  $(0,0)$, $(1/3,1)$ and  $(1/3,-1)$, respectively. 
Two different RF cells of POLARIS \cite{polaris}  with the working 
frequencies of 384.9 MHz and 320.1 MHz
have been used to provide the "2-6" and "3-5" transitions, respectively. 
The spin modes were cycled by spill-by-spill.  The polarized deuteron 
beam was accelerated up to $T_d^{lab}$=880~MeV by keeping the
quantization 
axis perpendicular to the beam-circulation plane of the Nuclotron. The 
typical beam intensity in the Nuclotron ring was 2--3$\times 10^7$ 
deuterons per spill with a duration of $\sim$1 s, independently on the
spin mode.

The detection system was designed for analyzing powers
measurements in a wide range of initial deuteron energies \cite{uesaka}.  
The detector support with mounted 46 plastic scintillation counters was
placed 
downstream the ITS spherical chamber. Each plastic scintillation
counter 
was coupled to a photo-multiplier tube Hamamatsu H7416MOD. 
Nine proton detectors were installed for left, right and up, but due
to 
space limitation -- only four for down directions.
The proton detectors were placed at a distance of 600 mm from the
target. The angular span of one proton detector was 2$^\circ$ in the 
laboratory frame, which corresponds to $\sim$4$^\circ$ in the c.m.
Four deuteron detectors  were placed at scattering
angles of deuterons coinciding kinematically with the protons for 
left, right and up scattering. Only one deuteron
detector was used to cover the solid angle corresponding to down scattering. 
In addition, one pair of detectors  was placed to register two protons 
from quasi-elastic $pp$- scattering at $\theta_{pp}$=90$^\circ$ in the
c.m. 
in the horizontal plane to monitor the beam luminosity. 
The deuteron- and quasi-elastic
$pp$- detectors were placed at a distance of 560 mm from the target in 
front of the proton detectors.

The scattered deuterons and recoil protons were detected in
kinematical 
coincidence 
over the center-of-mass  angular range of 60--140$^{\circ}$.
The analyzing powers $A_{y}$, $A_{yy}$ and $A_{xx}$ were measured at nine (eight)
different angles in the c.m. defined by the position of the proton 
counters placed in the 
horizontal (vertical) plane.

\begin{figure}[hbt]
\vspace{9pt}
\resizebox{0.48\textwidth}{!}  {
\includegraphics{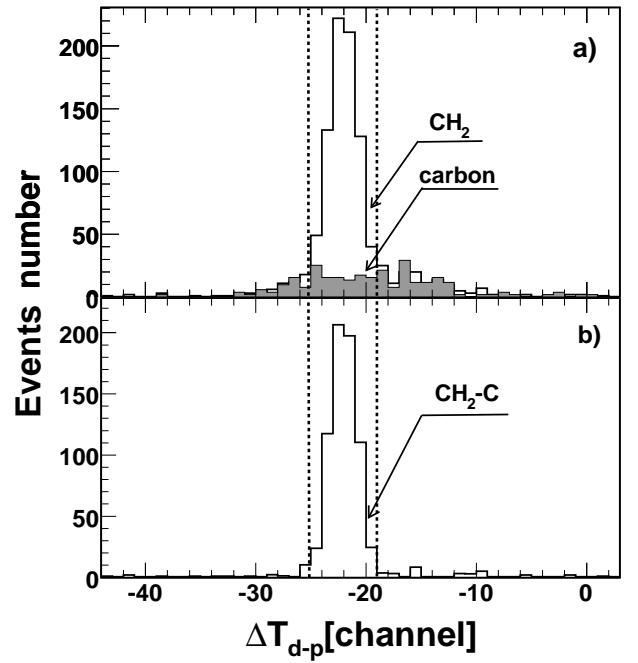}
}
\caption{The time difference between the signals for deuteron and
  proton 
detectors at $T_d^{lab}$=880~MeV and a scattering angle of 70$^\circ$
in 
the c.m. 
a) - the normalized spectra obtained for CH$_{2}$  and carbon targets
(open and shaded histograms, respectively); b) - the result of 
the CH$_2$-C subtraction.
The dashed lines are the prompt timing windows to select the $dp$-
elastic 
scattering events.
}
\label{fig:fig1}
\end{figure}

Selection of the $dp$- elastic scattering events was based on the 
kinematical coincidence of the scattered deuterons and recoil protons 
by the scintillation counters at several angles in the c.m. 
The energy loss correlation and  time-of-flight difference for the 
signals from the corresponding
proton-deuteron detector pairs were used.
The time difference between the signals from the deuteron and proton 
detectors for CH$_{2}$  and carbon targets 
gated by the correlation on the energy losses are shown in 
the upper panel in Fig.~\ref{fig:fig1} by the open and shadowed histograms,
respectively. The dashed lines in Fig.~\ref{fig:fig1}
represent the prompt timing window for the
selection of the $dp$- elastic scattering events.  
The carbon contribution is mostly due to quasi-free scattering and
appears 
as a broad bump. It is eliminated partly
by the criteria on the energy losses correlation in deuteron and
proton 
scintillation detectors. However, 
the contribution from the carbon content in CH$_2$ inside the timing window 
varied between 10 and 25\% depending on the detection angle.
The final selection of the $dp$- elastic events 
has been performed by the CH$_2$-C subtraction of the normalized 
time-of-flight difference spectra for each 
pair of the conjugated deuteron and proton detectors for each spin mode of
the PIS \cite{polaris}. The quality of the CH$_2$-C subtraction
is demonstrated in the bottom panel in Fig.~\ref{fig:fig1}.

The beam polarization has been measured before the data
taking using the asymmetry of the $dp$- elastic scattering  yields at 
$T_d^{lab}$=270~MeV
with the same detection system \cite{polar270}.  
Although the CH$_2$ film was used as a proton target, no measurements 
with a carbon target 
were made, because the background from the carbon content in
CH$_2$ was found negligibly small \cite{uesaka}, being in good
agreement with RIKEN measurements \cite{Sakamoto}. 
Data at several scattering angles were used to decrease the 
statistical errors of the beam polarization.
The values of the analyzing powers $A_y$, $A_{yy}$, $A_{xx}$ and $A_{xz}$ 
at these angles were obtained by the cubic spline interpolation of the
precise RIKEN data \cite{Sekiguchi02,suda}. 
The values of the vector $p_y$ and tensor $p_{yy}$ components of the beam
polarization for the "2-6" and "3-5" spin modes of POLARIS \cite{polaris} are
presented in Table~\ref{tbl:polarization}.
The systematical errors indicated in  Table~\ref{tbl:polarization} 
are due to the uncertainty of the $dp$- elastic scattering 
analyzing powers at $T_d^{lab}$=270~MeV \cite{Sekiguchi02,suda}.

\begin{table}[hbt]
\centering
\caption{The values of the vector $p_y$ and tensor $p_{yy}$ beam polarizations 
for the
"2-6" and "3-5" spin modes of POLARIS \cite{polaris}
obtained at $T_d^{lab}$=270~MeV \cite{polar270}.\label{tbl:polarization}}  
 
\begin{tabular}{ccccccc}
\hline\hline
 Spin & $p_y$ & $\Delta p_y^{stat}$ &  $\Delta p_y^{sys}$ &
$p_{yy}$ & $\Delta p_{yy}^{stat}$ &$\Delta p_{yy}^{sys}$ \\
  mode        & & & & & &\\
\hline
"2-6" & 0.216 & 0.014 & 0.002 & 0.605 & 0.024 & 0.005 \\
"3-5" & 0.208 & 0.011 & 0.002 & $-$0.575 & 0.020 & 0.005 \\
\hline\hline         
\end{tabular}
\end{table}

   The analyzing powers $A_y$, $A_{yy}$ and $A_{xx}$  for the  $dp$-
   elastic 
scattering at $T_d^{lab}$=880~MeV were measured simultaneously. 
The analyzing powers are defined in terms of the $xyz$ coordinate 
of $\vec{z}||\vec{k}_i$,  $\vec{y}||\vec{k}_i\times \vec{k}_f$, 
and $\vec{x}||\vec{y}\times\vec{z}$, 
where $\vec{k}_i$ and $\vec{k}_f$ are the incident and 
scattered deuteron momenta, respectively. 
The yields of the $dp$- elastic scattering with the vector $p_Z$ and 
tensor $p_{ZZ}$ deuteron polarizations
can be written as \cite{madison}
\begin{eqnarray}
\label{form:form1}
N_{pol}(\theta,\phi) &=& N_0(\theta,\phi)\cdot [ 1+\frac{3}{2} p_Z
  A_y(\theta) 
cos\phi \nonumber\\ 
&+&  \frac{1}{2} p_{ZZ} ( A_{yy}(\theta) cos^2\phi +A_{xx}(\theta) sin^2\phi )],
\end{eqnarray}
where $N_{pol}(\theta,\phi)$ and $N_0(\theta,\phi)$ are the yields
corrected by the beam luminosity and dead-time with 
polarized and unpolarized beams, respectively, $\theta$ is the scattering
angle,  and $\phi$ is the azimuthal angle with respect to the beam direction.
The azimuthal angles $\phi$ for the detectors placed in the 
directions of left, right, up, and down are $0$, $\pi$, $-\pi/2$, and 
$\pi/2$ radians, respectively. 
The analyzing powers $A_y$, $A_{yy}$ and $A_{xx}$ 
were extracted by using the normalized yields 
$n(\theta,\phi)= N_{pol}(\theta,\phi)/N_0(\theta,\phi)$, defined in Eq.~(\ref{form:form1})
\begin{eqnarray}
\label{form:form2}
A_y(\theta) &=& \frac{n(\theta,0)-n(\theta,\pi) }{3 p_Z}\nonumber\\
A_{yy}(\theta) &=& \frac{n(\theta,0)+n(\theta,\pi)-2}{p_{ZZ}}\\
A_{xx}(\theta) &=& \frac{n(\theta,-\pi/2)+n(\theta,\pi/2)-2}{p_{ZZ}}.\nonumber
\end{eqnarray}
Such a method does not require an accurate knowledge
of the detector geometries and/or efficiencies of the
detection system. The analyzing powers $A_y$, $A_{yy}$, and
$A_{xx}$ extracted by Eq.~(\ref{form:form2}) were averaged over 
"2-5" and "3-6" spin modes of POLARIS \cite{polaris}.

\section{Results and discussion}
\label{results}

\begin{figure}[hbt]
\vspace{9pt}
\resizebox{0.48\textwidth}{!}  {
\includegraphics{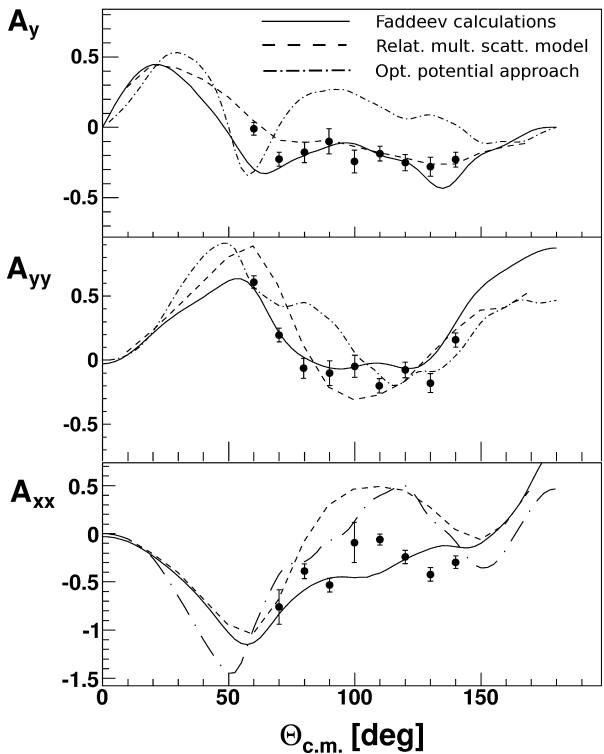}
}
\caption{The angular dependencies of the vector $A_y$ and 
tensor analyzing powers $A_{yy}$ and $A_{xx}$ for $dp$- elastic scattering 
at $T_d^{lab}$=880~MeV. The lines are explained in the text.
}
\label{fig:fig2}
\end{figure}

The angular dependencies of the vector $A_y$ and tensor analyzing powers $A_{yy}$ and
$A_{xx}$ obtained at $T_d^{lab}$=880~MeV are presented in  
Fig.~\ref{fig:fig2} by the solid symbols. The error bars are the 
statistical only.
The systematic errors for the vector $A_y$ and tensor analyzing powers $A_{yy}$ and
$A_{xx}$ from the uncertainty of the normalization of the
beam polarization are $\sim 7$\% and $\sim 4$\%, respectively. 
The systematic errors due to the procedure of the $dp$- elastic events 
selection were estimated to be 
$\sim 5.5$\% and 
$\sim 2.5$\% for vector and tensor analyzing powers, respectively.

The solid lines in Fig.~\ref{fig:fig2}
are the results of the nonrelativistic
Faddeev calculations \cite{CB1} using the CD-Bonn nucleon-nucleon 
potential \cite{cdbonn}. 
All partial waves with the total angular momentum of two-nucleon
subsystem  up to $j_{max}$=5 were taken into account. 
One can see that the results of Faddeev calculations based on CD-Bonn
potential even without involving 3NFs are 
in good agreement with all the analyzing powers within the achieved 
experimental accuracy.

The dashed lines correspond to a relativistic calculation 
in the multiple scattering expansion formalism \cite{nadia} 
up to the second-order terms of
the nucleon-nucleon $t$-matrix \cite{nadia_NN} with the use of 
the CD-Bonn \cite{cdbonn} DWF. The 
parameterization \cite{LF} of the $NN$ $t$-matrix \cite{nadia_NN} obtained from
the recent phase-shift analysis data SP07 \cite{PWA} allows to avoid the
convergence problem due to maximal number of partial wave states in the $NN$ system.
The model \cite{nadia} takes into account the off-energy-shell effects.
The approach describes reasonably well the angular dependencies of $A_y$ 
over the whole angular range of measurements  and  $A_{yy}$ at
backward angles. It fails to reproduce the behavior of  $A_{xx}$.  

The dot-dashed curves correspond to the relativistic 
calculation of the optical potential framework \cite{maxim} up to the total
angular momentum $J=39/2$. The results are obtained
with the use of DWF derived from the dressed bag model of the Moscow-T\"ubingen group \cite{kukulin} and the on-shell 
nucleon-nucleon $t$-matrix 
based on the recent phase-shift analysis data  SP07 \cite{PWA}. 
These calculations reproduce the behavior of  $A_{yy}$ only at 
the angles larger than 100$^\circ$ in the c.m., 
while they fail to describe the analyzing powers $A_y$ and $A_{xx}$.

\begin{figure}[hbt]
\vspace{9pt}
\resizebox{0.48\textwidth}{!}  {
\includegraphics{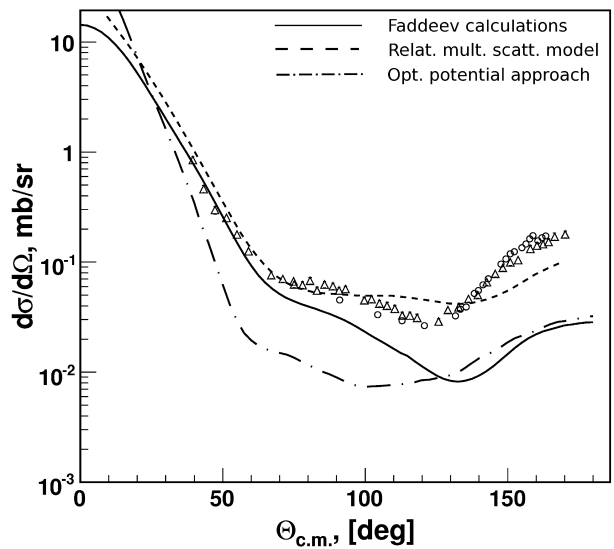}
}
\caption{The angular dependence of the $dp$- elastic scattering cross 
section at $T_d^{lab}\sim$880~MeV. The solid, dashed and dash-dotted 
curves are the results of the Faddeev calculation \cite{CB1}, of the 
relativistic multiple scattering model \cite{nadia,nadia1} 
and of the optical potential approach \cite{maxim}, respectively. 
The open triangles and circles correspond to
the data obtained at $T_d^{lab}$=850~MeV \cite{cs_1} 
and $T_d^{lab}$=940~MeV \cite{cs_2}, respectively.
}
\label{fig:fig3}
\end{figure}

The results of the nonrelativistic Faddeev calculations \cite{CB1}, 
the relativistic multiple scattering model \cite{nadia,nadia1} and 
the optical potential approach  \cite{maxim} for the $dp$- elastic 
scattering  differential cross section at $T_d^{lab}$=880~MeV are shown 
in Fig. \ref{fig:fig3} by the solid, dashed and dash-dotted curves, 
respectively. They are compared with the experimental
data  obtained at $T_d^{lab}$=850~MeV \cite{cs_1}  
and $T_d^{lab}$=940~MeV \cite{cs_2} given by the open triangles 
and circles, respectively.

The optical potential approach  \cite{maxim} which uses the DWF 
derived from the dressed bag model
\cite{kukulin} fails to reproduce the cross section data. Both 
non-relativistic Faddeev \cite{CB1} and relativistic multiple scattering 
\cite{nadia1} calculations describe the experimental data
up to the scattering angles in the c.m. of $\sim$70$^\circ$.  
The Faddeev approach \cite{CB1}
cannot reproduce the cross section data at larger angles,
while the relativistic
multiple scattering model calculations \cite{nadia1} provide better
agreement with the data in the vicinity of the
differential cross section minimum. 
Note that the relativistic effects in the framework of Faddeev 
calculations performed with NN forces are significant only 
in the region of backward angles
and they are small in the region of the differential cross  section 
minimum \cite{Faddeev_rel}. Therefore,  the relativistic effects 
alone cannot explain the difference between the predictions of these two models.
Both approaches cannot reproduce the data at backward scattering
angles, where  the essential contribution of the
$\Delta$-isobar excitation into the differential cross section at high 
energies was demonstrated \cite{kaptari}.

The agreement between the cross section 
data and theoretical calculations at the angles 
larger than 70$^\circ$ in the c.m. could be improved, if the
three-nucleon forces are taking into account. 
However, theoretical predictions including present 
3NF models \cite{Urbana9,TM99} or an effective 3NF due to 
explicit $\Delta$-isobar excitation \cite{Deltuva} 
underestimate the data at 250 MeV/nucleon \cite{Hatanaka02,Maeda07} 
by up to 40\%. 
The inclusion of present 3NFs in 
the calculations does not improve the description of the data on the
nucleon analyzing power and proton polarization transfer coefficients obtained
at 250 MeV \cite{Hatanaka02,Maeda07}.
Therefore, at higher
energies, the deviation of both  spin observables and cross
sections 
from the theoretical calculations 
\cite{CB1,Deltuva,Faddeev_rel}
demonstrates the deficiencies of the present 3NF models and
relativistic 
description of the
$dp$ -elastic scattering process. 
This might indicate that additional
short-range 3N forces should be added to the 2$\pi$ -exchange type 
forces \cite{Faddeev_rel,Faddeev_rel1}.

The relativistic multiple scattering model 
describes reasonably well the $dp$- elastic scattering 
differential cross section and vector analyzing power  $A_y$
up to 140$^\circ$ in the c.m. \cite{nadia,nadia1}. However,
it reproduces the behavior of the tensor analyzing power $A_{yy}$ 
only at backward angles, while the angular dependence of $A_{xx}$ 
is not described. The agreement between the experimental data and theoretical description could be improved, if the 3NF are included into consideration. 

\section{Conclusions}
\label{conclusions}

The vector $A_y$ and tensor analyzing powers $A_{yy}$ and $A_{xx}$ for $dp$- elastic scattering have been measured 
for the first time at Internal Target Station at the JINR Nuclotron at $T_d^{lab}$=880~MeV 
over the c.m. angular range from 60$^\circ$ to 140$^\circ$ {corresponding to
the transverse momenta of $\sim$400-600~MeV/$c$}.

{New results on the tensor analyzing powers indicate strong deviations from  
the predictions of the relativistic phenomenological approaches  
\cite{nadia, maxim} based on the use of the nucleon-nucleon forces only.
The non-relativistic  Faddeev calculations \cite{CB1} using CD-Bonn 
2NF \cite{cdbonn} describe only the angular 
behavior of the new data for the analyzing powers. They, however, fail to  
reproduce 
the differential cross section data obtained
 in earlier experiments \cite{cs_1,cs_2}.}  
 
Some deficiencies in the description of the differential
cross section and the deuteron analyzing powers at $T_d^{lab}\sim$880~MeV
obtained at quite large transverse momenta 
require the consideration of additional mechanisms, for 
instance, 3NFs. 
Since present 3NFs models cannot improve the agreement with the data obtained
at lower energies \cite{Hatanaka02,Maeda07}, 
new models of 3NFs (including short-range part) 
should be considered.

\vskip 5mm
The authors are grateful to the Nuclotron accelerator  and POLARIS groups. 
They thank L.S.~Azhgirey, 
Yu.S.~Anisimov,  E.~Ayush, A.F.~Elishev, V.I.~Ivanov,
L.V.~Karnjushina, 
J.~Kliman,  Z.P.~Kuznezova, A.P.~Laricheva, A.G.~Litvinenko,
V.~Matousek, 
M.~Morhach,
V.G.~Perevozchikov, V.M.~Slepnev, I.~Turzo,  Yu.V.~Zanevsky and 
V.N.~Zhmyrov  for their help during the preparation and performance of the 
experiment. 
The investigation has been  partly supported 
by the Grant-in-Aid for Scientific Research (Grant No. 14740151) of 
the Ministry of Education, Culture, Sports, Science, and Technology of Japan; 
by the Russian Foundation for 
Fundamental Research (Grant No. 10-02-00087-a); 
and by the Polish National Science Center as research project
No. DEC-2011/01/B/ST2/00578. Part of the numerical calculations were
performed on the supercomputer cluster of the JSC, J\"ulich, Germany.








\end{document}